\documentclass[
aps,nofootinbib,showpacs,showkeys,preprint
,tightenlines ] {revtex4}

\usepackage{epsf,epsfig,subfigure,graphicx,amsmath,amssymb}
\usepackage{color}

\newcommand{\dis}[1]{\begin{equation}\begin{split}#1\end{split}\end{equation}}
\newcommand{\be}{\begin{equation}}
\newcommand{\ee}{\end{equation}}

\newcommand{\eq}[1]{Eq.~(\ref{#1})}

\def\bea{\begin{eqnarray}}
\def\eea{\end{eqnarray}}

\begin{document}

 \begin{flushright}
{
}
\end{flushright}

\title{\Large\bf Singlet Extension of the MSSM
\\
for 125 GeV Higgs Mass with the Least Tuning
}

\author{
Bumseok Kyae$^{(a)}$\footnote{email: bkyae@pusan.ac.kr}
and Jong-Chul Park$^{(b)}$\footnote{email: jcpark@kias.re.kr}
}
\affiliation{$^{(a)}$
Department of Physics, Pusan National University, Busan 609-735, Korea
\\
$^{(b)}$ Korea Institute for Advanced Study, Seoul 130-722, Korea
}

\begin{abstract}

In order to raise the Higgs mass to 125 GeV and relieve the
fine-tuning associated with the heavy s-top mass in the minimal supersymmetric standard model (MSSM), we
propose a new singlet extension of the MSSM. In this scenario,
the additional Higgs mass is radiatively generated in a hidden sector,
and the effect is transmitted to the Higgs through a messenger
field. The Higgs mass can be efficiently increased by the
parameters of the superpotential as in the extra matter scenario,
but free from the constraints on extra colored matter fields by
the LHC experiments.
As a result, the tuning problem can be remarkably mitigated by
taking low enough messenger mass ($\sim 300$ GeV) and mass
parameter scales ($\sim 500$ GeV). We also discuss how to enhance
the diphoton decay rate of the Higgs over the SM expectation in
this framework.
\end{abstract}


\maketitle


\section{Introduction} \label{sec1}

Recently, CMS and ATLAS reported the observations of the signals,
which can be interpreted as the presence of a standard model
Higgs(-like) boson with the mass of 125 GeV around the five sigma
confidence level \cite{CMS,ATLAS}.  The news seems to be accepted
as the discovery of the long-awaited Higgs particle, which is very
essential in mass generations for the standard model (SM)
particles.
%
%
%
However, the theoretical issues associated with the Higgs boson,
e.g. how the Higgs can naturally exist at low energies, still
remain unsolved. Actually, these issues have played the role of
strong motivations to study various new physics beyond the SM.

For the last three decades, the minimal supersymmetric standard
model (MSSM) has maintained the status as the leading candidate
beyond the SM \cite{MSSM}. The MSSM provides a beautiful solution
to the large hierarchy problem between the electroweak(EW) and the
grand unification (GUT) or Planck scales with the minimal
extension of the SM in the supersymmetric (SUSY) way, which makes
it possible to embed the SM in a fundamental theory like the
string theory \cite{string}. The gauge coupling unification is
another great advantage of the MSSM.

\vspace{0.3cm} \underline{\bf MSSM}: In the MSSM, a relatively
smaller Higgs mass is preferred.
It is basically because the tree-level quartic coupling of the
Higgs potential is given by the small gauge coupling unlike the
SM. As a result, the Higgs mass cannot be even larger than the $Z$
boson mass ($M_Z$) without large radiative corrections: by
including the radiative correction by the large top quark Yukawa
coupling, the Higgs mass can be lifted above 100 GeV.
%
Actually, the lightest Higgs mass in the MSSM is given by
\dis{ \label{mh2MSSM}
m_h^2
&\approx M_Z^2{\rm cos}^22\beta+\frac{3}{4\pi^2}\left(y_tm_t\right)^2{\rm sin}^2\beta~{\rm log}\left(\frac{m_t^2+\widetilde{m}_t^2}{m_t^2}\right) ,
} where $y_t$ is the top quark Yukawa coupling, and $m_t^2$ and
$\widetilde{m}_t^2$ denote the mass squared of the top quark and
the soft mass squared of its superpartner ``s-top,'' respectively.
The factor ``3'' results from the number of the colors which the
(s-)top carries. The first term on the right-hand side comes from
the tree-level contribution, and the second term from the
radiative correction ($\equiv\Delta m_h^2|_{\rm MSSM}$) by the top
quark and the s-top. Here we neglect the ``$A$-term''
contribution. In \eq{mh2MSSM}, however, the values of the top
quark mass $m_t$ and also the top quark Yukawa coupling $y_t$ (up to ${\rm tan}\beta$) have
already been precisely measured. Thus, the only useful parameter
for raising the Higgs mass is the soft mass squared of the s-top,
$\widetilde{m}_t^2$. Note that the radiative correction {\it
logarithmically} depends on the s-quark mass squared
($m_t^2+\widetilde{m}_t^2$). Thus, raising the Higgs mass with the
soft mass squared of the s-top is not a quite efficient way.
Indeed, an s-top mass larger than a few TeV is needed to achieve
125 GeV Higgs mass at two-loop level, unless the large mixing
effect between the left and right s-tops through the $A$-term
contribution is assumed \cite{MSSM,twoloop}.
However, the s-top mass cannot be arbitrarily large.

The radiative correction by the top and s-top also contributes to
the renormalization of the soft parameter $m_2^2$, which is the
soft mass squared of the MSSM Higgs, $H_u$: \dis{ \label{m2MSSM}
m_{2}^2(M_Z)\approx m_{2}^2-\frac{3y_t^2}{8\pi^2}
\widetilde{m}_t^2 ~{\rm
log}\left(\frac{M_G^2}{\widetilde{m}_t^2}\right) ,
 }
where $M_G$ indicates the GUT scale ($\approx 2\times 10^{16}$
GeV), at which the soft parameters are assumed to be generated in
the minimal supergravity (SUGRA) model. Here we keep only the radiative
correction coming from the top quark Yukawa coupling, which is the
largest correction to $m_2^2$. The negative contribution of the
last term in \eq{m2MSSM} causes the sign flipping of $m_2^2$ at
the EW energy scale, which triggers the EW symmetry breaking.
Thus, one of the extremum conditions for the MSSM Higgs fields is
modified as \dis{ \label{m22MSSM} m_{2}^2 + |\mu|^2
\approx  m_3^2{\rm cot}\beta + \frac{M_Z^2}{2}{\rm cos}2\beta -
\frac{3y_t^2}{8\pi^2} \widetilde{m}_t^2 ~{\rm
log}\left(\frac{\widetilde{m}_t^2}{M_G^2}\right) . } The radiative
corrections add the last term ($\equiv\Delta m_2^2$) in
\eq{m22MSSM}. If a too heavy s-top mass is taken to raise the
Higgs mass by \eq{mh2MSSM}, $\Delta m_2^2$ and other parameters
should be properly tuned to give $M_Z^2$, which implies that the
EW symmetry breaking becomes unnatural. Actually, \eq{m22MSSM} is
not directly related to the observed value of the Higgs mass but
closely associated with the naturalness of the EW symmetry
breaking. It is known as the ``little hierarchy problem'' in the
MSSM. Thus, e.g. for  $\widetilde{m}_t$ of 2 TeV, the size of the
tuning is roughly estimated by the hierarchy in the relation of
\eq{m22MSSM}: \dis{ \frac{(M_Z^2/2){\rm cos}2\beta}{|\Delta
m_{2}^2|} <\left| \left(\frac{\widetilde{m}_t^2}{M_Z^2}\right)
\frac{3y_t^2}{4\pi^2} ~{\rm
log}\left(\frac{\widetilde{m}_t^2}{M_{G}^2}\right)
\right|^{-1}\lesssim 4.7\times 10^{-4}  . } In order to reduce the
tuning in \eq{m22MSSM}, thus,

\vspace{0.3cm}

$\bullet$ smaller mass parameters need to be taken, but yielding
$m_h=125$ GeV;

$\bullet$ a low energy soft term generation scenario  is needed
for a smaller log piece in \eq{m22MSSM}.

\vspace{0.3cm} \noindent In this paper, we will introduce a
phenomenologically attractive scenario, addressing the above two
requirements.

\vspace{0.3cm} \underline{\bf Maximal Mixing}: In fact, 125 GeV
Higgs mass could be achieved even with relatively lighter s-tops
by considering also the ``$A$-term'' contribution to the radiative
correction, which was dropped in \eq{mh2MSSM}. A large mixing
between the s-tops of the SU(2)$_L$ doublet and singlet,
$(\tilde{t}_L, \tilde{t}_R)$, via the SUSY breaking ``$A$-term''
is very helpful for raising the Higgs mass. Particularly, the
``maximal mixing'' \dis{ X_t\equiv (A_t-\mu{\rm
cot}\beta)=\sqrt{6}~ m_{\tilde{t}} ~, } where
$m_{\tilde{t}}\equiv\sqrt{m_t^2+\widetilde{m}_t^2}$, can lift the
Higgs mass up to 135 GeV without any other helps in the decoupling
limit of the CP odd Higgs \cite{MSSM}. However, as the mixing
deviates from the maximal mixing, the enhancement effect drops
rapidly. Employing a large mixing of $\tilde{t}_L$-$\tilde{t}_R$,
hence, would be a kind of fine-tuning in this sense. Throughout
this paper, we will not consider such a mixing effect.

%

\vspace{0.3cm} \underline{\bf Extra Matter}: In order to
efficiently enhance the radiative correction, one might introduce
the fourth family of chiral matter or extra vectorlike matter
\cite{extramatt,moroi}. In the case of the fourth family of the
chiral matter, the top quark Yukawa coupling and also the top
quark mass in \eq{mh2MSSM} are replaced by the unknown parameters,
which can be utilized to enhance the Higgs mass. Since such SUSY
parameters appear outside the logarithmic function, they can
efficiently increase the Higgs mass unlike the s-top mass squared
in the MSSM. However, the presence of extra colored particles
{\it coupled to the Higgs} with order-one Yukawa couplings would
exceedingly affect the production rate and also decay rate of the
Higgs at the large hadron collider (LHC), i.e. $gg\rightarrow h$
and  $h\rightarrow\gamma\gamma$: they result in immoderate
deviation from the LHC data. According to Ref.~\cite{4thfamily},
indeed, the existence of such an extra family of the chiral matter
is excluded at the 99.9$\%$ confidence level for the Higgs mass of
125 GeV.

In the case of extra vectorlike matter, in which  a Yukawa
coupling of order unity with the Higgs is  still necessary for
lifting the Higgs mass, the LHC bound could be avoided by
employing heavy enough mass terms for vectorlike fields. However,
the tuning problem associated with the naturalness of the Higgs
mass becomes serious with the high scale mass
parameters.\footnote{For instance, if only an extra vectorlike
pair of quark doublets $\{Q,Q^c\}$ is introduced and the
superpotential $W=M_QQQ^c+yQH_uu^c$, where $H_u$ and $u^c$ are the
Higgs and a quark singlet in the MSSM, is considered, using the
formula in \cite{moroi} one can show that the radiative correction
to the Higgs potential is \dis{ \Delta
V=\frac{3}{16\pi^2}\left[\left(M^2+\widetilde{m}^2\right)^2\left\{{\rm
log}\left(\frac{M^2+\widetilde{m}^2}{\Lambda^2}\right)-\frac{3}{2}\right\}-M^4\left\{{\rm
log}\left(\frac{M^2}{\Lambda^2}\right)-\frac{3}{2}\right\}\right]+{\rm
constant} , } where $M^2\equiv M_Q^2+y^2|H_u|^2$ and $\Lambda$
indicates a renormalization scale. Here all the soft mass squareds
are set to be $\widetilde{m}^2$, and the ``$A$-term'' effect is
ignored for simplicity. This expression is quite similar to that
in the case of Ref.~\cite{KP}. However, the fields circulating on
the loops in Ref.~\cite{KP} are MSSM singlets.} Moreover, the
extra vectorlike matter should compose the SU(5) or SO(10)
multiplets to protect the gauge coupling unification. If the low
energy effective theory is not embedded in four-dimensional SU(5)
or SO(10) GUTs but in other unified theory defined in higher
dimensional spacetime like string theory \cite{string}, we need to
explore other possibilities to explain the 125 GeV Higgs mass.

\vspace{0.3cm} \underline{\bf NMSSM}: In the next-to-minimal
supersymmetric standard model (NMSSM), the Higgs mass can be
raised by the tree-level correction of the Higgs potential
\cite{nmssm,nmssm2,nmssm3}. In the NMSSM, the MSSM $\mu$ term is
promoted to a renormalizable trilinear term $SH_uH_d$ in the
superpotential, introducing an extra singlet superfield $S$
together with a dimensionless coupling $\lambda$. The presence of
such a trilinear term in the superpotential provides the quartic
coupling to the Higgs potential as well as a solution to the $\mu$
problem through the gravity mediated SUSY breaking scenario.
By the quartic Higgs potential coming from $\lambda SH_uH_d$ in
the superpotential, the mass of the lighter CP even Higgs in the
NMSSM is modified at the tree-level as \dis{ \label{HmassNMSSM}
m_h^2\approx M_Z^2{\rm cos}^22\beta+\lambda^2v_H^2{\rm
sin}^22\beta + \Delta m_h^2|_{\rm MSSM} ~, } where $v_H^2\equiv
v_u^2+v_d^2=(174~{\rm GeV})^2$ and $\Delta m_h^2|_{\rm MSSM}$
denotes the radiative correction by the (s-)top.
The tree-level correction ``$\lambda^2v_H^2{\rm sin}^22\beta$'' in
\eq{HmassNMSSM} can remarkably raise the Higgs mass, if the
dimensionless Yukawa coupling $\lambda$ is sizable. In order to
maintain the perturbativity of the model up to the GUT scale,
however, $\lambda$ is known to be  smaller than $0.7$ at the EW
scale (``Landau pole constraint'') \cite{nmssm}. Moreover, to
achieve the Higgs mass of 125 GeV with the s-top mass much lighter
than 1 TeV, which is necessary for the naturalness of the Higgs,
$\lambda$ needs to be larger than $0.5$. Requiring both the
perturbativity and the naturalness, thus, the allowed range of
$\lambda$ should be quite limited: \dis{  \label{lambdaNMSSM}
0.5\lesssim \lambda \lesssim 0.7. } The relatively small $\lambda$
pushes ${\rm tan}\beta$ to the smaller values for the 125 GeV
Higgs mass: \dis{ \label{tbetaNMSSM} 1\lesssim {\rm
tan}\beta\lesssim 3 , } which gives almost the maximal values to
${\rm sin}^22\beta$ in \eq{HmassNMSSM}.


%

\vspace{0.3cm} \underline{\bf Radiative Correction by MSSM
Singlets}: Recently, the authors of Ref.~\cite{KP} proposed a
scenario in which the Higgs mass is raised through {\it radiative
corrections by some MSSM singlet fields}. In this case, the Higgs
mass can be efficiently lifted by using the  parameters of the
superpotential just like the extra matter case, but the LHC
constraint can be avoided because only MSSM singlets are employed.
In Ref.~\cite{KP}, it was shown that the parameter space of ${\rm
tan}\beta$ and the trilinear coupling of ``$SH_uH_d$'' ($\equiv
y_H$) in the superpotential to explain the 125 GeV Higgs mass can
be remarkably enlarged by extending the NMSSM with some other MSSM
singlets, compared to the original form of the NMSSM: $0.2\lesssim
y_H\lesssim 0.5$ and $3\lesssim {\rm tan}\beta\lesssim 10$ can be
also consistent with the Higgs mass of 125 GeV even without the
mixing effect.

Since the Higgs mass is radiatively generated from a hidden sector
and then it is transmitted to the Higgs sector through a mediation
by a messenger in this scenario, the fine-tuning problem can be
quite alleviated by taking low scale messenger and mass
parameters. In this paper, we will particularly discuss how much
the fine-tuning in the Higgs sector can be relieved in this setup.

\vspace{0.3cm} This paper is organized as follows. In section
\ref{sec2}, our basic setup will be introduced. In section
\ref{sec3}, the effective Higgs potential will be calculated in
our setup. In section \ref{sec4}, we will discuss how to achieve
the 125 GeV Higgs mass and minimize the tuning. In section
\ref{sec5}, we will briefly discuss how to enhance the diphoton
decay rate of the Higgs in our framework. In section \ref{sec6},
we will propose a UV model. Section \ref{sec7} will be devoted to
the conclusion.


\section{A Singlet Extension of the MSSM} \label{sec2}

In this paper, we will pursue the naturalness of the model rather
than its minimality. Introducing the {\it MSSM singlet}
superfields $\{S,S^c\}$ and $\{N,N^c\}$, we extend the MSSM Higgs
sector in the superpotential as follows: \dis{ \label{superPot}
W=\left(\mu + y_HS\right)H_uH_d + \mu_SSS^c + \left(\mu_N +
y_NS^c\right)NN^c ,} where $\{H_u, H_d\}$ denote the two MSSM
Higgs doublets.\footnote{ If we should seriously accept the
recently observed excess of the diphoton decay rate of the Higgs
\cite{CMS, ATLAS}, we need to slightly modify this model. In
section \ref{sec5}, we will assign also electromagnetic charges to
$\{N,N^c\}$ just for the explanation of the excess under the
assumption that the diphoton decay rate of the Higgs will not
approach to the SM prediction even with more data. In other
sections, however, we will ignore the diphoton excess and so
regard $\{N,N^c\}$ as neutral fields under the SM.} For
simplicity, we assume that the parameters in \eq{superPot} are all
real. Since the $\mu$ and $\mu_{S,N}$ terms are explicitly
present, there remains no Pecci-Quinn (PQ) symmetry at the EW
scale. Apart from the MSSM $\mu$ term, the trilinear term
$SH_uH_d$ {\it a la} the NMSSM is introduced in \eq{superPot}
\cite{KNS,JSY}. 

Equation (\ref{superPot}) should be regarded as a low energy effective superpotential, 
which is embedded in a UV superpotential with more (global) symmetries. 
As a result of symmetry breaking in the UV theory, \eq{superPot} can be deduced. 
Otherwise, including the tadpole terms of the singlets $S$ and $S^c$, all the powers of them had to appear in the superpotential for the consistency, 
since $S$ and $S^c$ cannot carry any quantum numbers only with \eq{superPot}. 
Moreover, a gauge- and global-symmetry singlet is known to destabilize the gauge hierarchy, 
provided it has renormalizable couplings to the visible fields \cite{tadpole1,tadpole2}. 
How \eq{superPot} can be generated from a UV superpotential, under which the singlets $S$, $S^c$ carry (global) charges, will be discussed in section \ref{sec6}.

$\{S,S^c\}$ are the messenger fields, which connect  the Higgs
$\{H_u,H_d\}$ and the hidden sector fields $\{N,N^c\}$. Note that
the ``messenger'' and ``hidden sector'' here do not necessarily
mean the conventional ones appearing in various SUSY breaking scenarios.
The hidden gauge interaction is not confining here: it is assumed to remain perturbative down to the EW scale. We only require the mass splitting between the bosonic and fermionic modes in the hidden sector superfields $\{N,N^c\}$ such that they eventually generate the radiative correction of the Higgs mass. Such an effect can be
transmitted to the Higgs via the messengers
$\{S,S^c\}$ as will be seen later.
$\{N,N^c\}$ form a vectorlike $n$-dimensional representation of a
certain hidden gauge group. They could remain light down to low
energies due to global symmetries.

$\mu_{S,N}$ terms are the Dirac type bare mass terms of the
messengers and hidden sector fields. $\mu_{S,N}$ are assumed to be
larger than 300 GeV. Thus, the squared masses of
$\{\widetilde{S},\widetilde{S}^c\}$ and
$\{\widetilde{N},\widetilde{N}^c\}$, which are the scalar
components of $\{S,S^c\}$ and $\{N,N^c\}$, respectively, are quite
heavier than that of the lightest Higgs. Since $\mu_S$ and $\mu_N$
both are much larger than the Higgs mass, there is no
``singlet-ino'' (the fermionic components of singlet superfields)
lighter than the Higgs. Thus, there is no invisible decay channel
of the Higgs in this model. However, we restrict $\mu_{S,N}$ to be
smaller than 1 TeV. It is because the fine-tuning in the Higgs
sector would become serious if they are heavier than 1 TeV. Their
smallness compared to the fundamental scale will be explained in
section \ref{sec6}.

In fact, the superpotential \eq{superPot} can provide a quartic
Higgs potential at the tree-level as in the NMSSM, which is quite
helpful for lifting the Higgs mass if $y_H$ can be sizable.
However, the Landau pole constraint to avoid the blow-up of $y_H$
below the GUT scale is known to restrict the size of $y_H$ to be
smaller than 0.7 \cite{nmssm}. While $y_H$ should be smaller than unity, $y_N$, which is the Yukawa coupling of $S^cNN^c$ in
\eq{superPot}, {\it can still be of order unity} at the EW scale.
Nonetheless, the hidden gauge interaction of $\{N, N^c\}$ can
prevent $y_N$ from the blow-up at higher energy scales, because
$\{N, N^c\}$ carry a non-Abelian gauge charge of a relatively
large hidden gauge group.

For instance, if the hidden gauge group is SU(5)$_H$, under which
$\{N,N^c\}$ are $n=5$ representations, and the beta function
coefficient $b_H$ is $-4$, $y_N$ smaller than 2.3 at the EW scale
still decreases with energy up to the GUT scale, assuming that the
gauge coupling of the hidden gauge group, $g_H$ is unified with
the visible sector gauge couplings at the GUT scale.\footnote{The
renormalization group equations of the hidden gauge coupling $g$
and $y_N$ are \dis{ 16\pi^2\frac{dg_H}{dt}=b_Hg_H^3 ,\quad
16\pi^2\frac{dy_N}{dt}=y_N\left[(n+2)y_N^2-4g_H^2C_2(F)\right] , }
where $t$ parametrizes the energy scale, $t_0-t={\rm
log}(\Lambda_{\rm UV}/\mu)$. For SU$(N)_H$ hidden gauge group, the
beta function coefficient $b_H$ ($=-3N+\sum T_R$) is determined by
matter contents of the hidden sector. $C_2(F)$ for the 
fundamental representation of the SU$(N)_H$ generators,
$(T^aT^a)_i^j=C_2(F)\delta_i^j$, is given by $C_2(F)=(N^2-1)/2N$.
} In this case, $\alpha_H$ ($\equiv g_H^2/4\pi$) at the EW scale
($\approx 0.2$) is still in the perturbative regime. If SU(5)$_H$ is
embedded in other groups or more matter fields can be relevant above the intermediate scale, we have more possibilities. SU(5)$_H$
should be eventually broken or confining, but it is not much
important here only if the breaking scale is low enough.

Since $y_H$ is relatively small and $\mu_S$ is quite heavier than
the Higgs mass, the tree-level correction by $\{S, S^c\}$ to the
Higgs potential is expected to be suppressed. Moreover, the mixing
angles between the Higgs and the singlet sectors would be
negligible. In Ref.~\cite{KP}, however, it was shown that even
with relatively small $y_H$ ($0.2$-$0.5$), the Higgs mass of 125
GeV can be achieved through the large radiative correction if a
relatively larger $y_N$ compensates the smallness of $y_H$.

With small enough $y_H$ the soft mass squared of $S$,
$\widetilde{m}_S^2$ does not run much with energy at one-loop
level. On the other hand, $y_N$ is of order unity, and so
$\widetilde{m}_{S^c}^2$ can be suppressed at low energies compared
to $\widetilde{m}_S^2$ by the renormalization group (RG) effect.
Due to the gauge interaction in the hidden sector, the soft masses
of $N$ and $N^c$, $\widetilde{m}_{N}$ and $\widetilde{m}_{N^c}$
can be quite heavier than other soft masses at low energies. For
simplicity of the future calculation, but considering the RG
behaviors, we assume a hierarchy among the mass parameters at low energies (below the scale of $\mu_S$):
\dis{
\label{hierarchy} \widetilde{m}_{S^c} ~\lesssim~  \mu ~\lesssim~
m_{3/2}, ~\mu_S ~\lesssim~ \widetilde{m}_{S} ~\lesssim~ \mu_N,
~\widetilde{m}_N~(=\widetilde{m}_{N^c}) ,} where $m_{3/2}$
collectively denotes typical soft parameters except
$\widetilde{m}_S$ and $\widetilde{m}_{S^c}$. Although
$\widetilde{m}_{S^c}$ is the smallest, the scalar component of
$S^c$ is still much heavier than the Higgs because its physical
mass squared is given by $\mu_S^2+\widetilde{m}_{S^c}^2$.
%


%

%

%


%
%
%
%
%


\section{The Effective Higgs Potential} \label{sec3}

Let us first integrate out the quantum fluctuations of
$\{N,N^c\}$. Due to the mass difference between the bosonic and
fermionic components in $\{N,N^c\}$, the one-loop effective
potential of $\widetilde{S}^c$ is generated \cite{CW}:
\begin{eqnarray} \label{1-loopPot}
\Delta V=\frac{n}{16\pi^2}\bigg[\left(M_N^2+\widetilde{m}_{N}^2\right)^2\left\{{\rm
log}\left(\frac{M_N^2+\widetilde{m}_{N}^2}{\Lambda^2}\right)-\frac32\right\}
 -M_N^4\left\{{\rm
log}\left(\frac{M_N^2}{\Lambda^2}\right)-\frac32\right\} \bigg] ,
\end{eqnarray}
where $\Lambda$ denotes a renormalization mass scale. 
As will be discussed later, $\Lambda$ will be chosen to be $\mu_S$, which is about one half of $\mu_N$ in our case, since all the extra singlets $\{S,S^c\}$ and $\{N,N^c\}$ introduced for enhancing the radiative correction to the Higgs mass are decoupled below the $\mu_S$ scale. 
The SUSY
mass of $\{N,N^c\}$ ($\equiv M_N$) is given by the summation of
$\mu_N$ and the classical value of $\widetilde{S}^c$ as explicitly
seen in the superpotential \eq{superPot}, and so
\dis{ \label{M_NSc} 
M_N^2=\left|\mu_N+y_N\widetilde{S}^c\right|^2
}
Thus, $\Delta V$ in \eq{1-loopPot} depends only on  $\widetilde{S}^c$.
Note that the hidden gauge sector is not involved in generating the effective potential of $\widetilde{S}^c$ at one-loop level, \eq{1-loopPot}.


Including the soft terms and the one-loop effective potential
obtained after integrating out $\{N, N^c\}$, $\Delta
V(\widetilde{S}^c)$, the scalar potential associated with the
superpotential \eq{superPot} is derived as follows:
\dis{
\label{hsPot} &V_{\rm
HS}=\left(m_2^2+|\mu+y_H\widetilde{S}|^2\right)|H_u|^2
+\left(m_1^2+|\mu+y_H\widetilde{S}|^2\right)|H_d|^2 \\
&\quad~~ +\left(\widetilde{m}_{S^c}^2+\mu_S^2\right)|\widetilde{S}^c|^2 +\left(\widetilde{m}_{S}^2+\mu_S^2\right)|\widetilde{S}|^2 + y_H^2|H_uH_d|^2  \\
&+\left[\left(y_H\mu_S\widetilde{S}^{c*}
+B_\mu\mu+y_HA_S\widetilde{S}\right)H_uH_d + B_S\mu_S\widetilde{S}\widetilde{S}^c + {\rm h.c.}\right] \\
&~ +\frac{1}{8}(g^2+g^{\prime
2})\left(|H_u|^2-|H_d|^2\right)^2+\frac12g^2|H_u^\dagger H_d|^2
+\Delta V(\widetilde{S}^c) ~, } where $B_\mu$, $B_S$, and $A_S$
denote the soft SUSY breaking ``$B$'' and ``$A$'' parameters. Here
we set $\widetilde{N}=\widetilde{N}^c=0$ for such heavy scalars, which fulfill all the
extremum conditions of the scalar potential.

Now let us integrate out $\{\widetilde{S},\widetilde{S}^c\}$,
which are heavier than $\{H_u,H_d\}$. The equations of motion in
the static limit for $\{\widetilde{S},\widetilde{S}^c\}$ are
\begin{eqnarray} \label{extremum}
&&\qquad\qquad\quad
\frac{\partial V_{\rm HS}}{\partial\widetilde{S}^c}=
\left(\widetilde{m}_{S^c}^2+\mu_S^2\right)\widetilde{S}^{c*}
+B_S\mu_S\tilde{S}+y_H\mu_SH_u^*H_d^*
+\partial_{\widetilde{S}^c}\Delta V = 0 ,
\\
&&\frac{\partial V_{\rm HS}}{\partial\widetilde{S}}=
\left(\widetilde{m}_{S}^2+\mu_S^2\right)\widetilde{S}^{*}
+B_S\mu_S\tilde{S}^c+y_HA_SH_uH_d
+\left(y_H\mu+y_H^2\tilde{S}^*\right)
\left(|H_u|^2+|H_d|^2\right)=0 .
\nonumber
\end{eqnarray}
Considering the hierarchy suggested in \eq{hierarchy}, the
approximate solutions to \eq{extremum} are given by
\dis{
\label{ScS} &~\widetilde{S}^c\approx
\frac{-1}{\mu_S^{2}}\bigg[y_H\mu_
SH_uH_d(1+\epsilon_1-\epsilon_2^*)
+\partial_{\widetilde{S}^{c*}}\Delta V^*(1+\epsilon_1)\bigg] ,
\\
&\widetilde{S}\approx \frac{-1}{\widetilde{m}_S^{2}+\mu_S^2}
\left[y_H\left(A_S^*-B_S^*\right)H_u^*H_d^*-\frac{B_
S^*}{\mu_S}\partial_{\widetilde{S}^c}\Delta V\right] \ll
\widetilde{S}^c , }
where
the terms proportional to $\widetilde{m}_{S^c}$ and $\mu$ are
ignored due to their relative smallness in \eq{hierarchy}, and
$\epsilon_1$ and $\epsilon_2$ are defined as \dis{
\epsilon_1\equiv \frac{|B_S|^2}{\widetilde{m}_S^{2}+\mu_S^2}
~~~{\rm and}~~~ \epsilon_2\equiv
\frac{A_S^*B_S}{\widetilde{m}_S^{2}+\mu_S^2} ~, } respectively.
Inserting the expressions of the heavy fields in \eq{ScS} into the
scalar potential $V_{\rm HS}$ of \eq{hsPot}, one can obtain the
low energy effective Higgs potential: \dis{ \label{effPot} V_{\rm
H}\approx&\left(m_2^2+\mu^2\right)|H_u|^2
+\left(m_1^2+\mu^2\right)|H_d|^2 + \left( B_\mu\mu H_uH_d+{\rm h.c.}\right)  \\
&~~~ +\frac{1}{8}(g^2+g^{\prime 2})\left(|H_u|^2-|H_d|^2\right)^2+\frac12g^2|H_u^\dagger H_d|^2 \\
&
~~+\left(\frac{\widetilde{m}_{S^c}^2}{\mu_S^2}-\frac{|A_S-B_S|^2}{\widetilde{m}_S^{2}+\mu_S^2}\right)y_H^2|H_uH_d|^2+\Delta
V(H) , } which is valid {\it below the mass scale of
$\{\widetilde{S},\widetilde{S}^c\}$.} Here we dropped the 
two-loop effects coming from
$|\partial_{\widetilde{S}^c}\Delta V|^2$. Since  $\partial_{\widetilde{S}^c}\Delta V$ is of order one-loop, only the
first term in \eq{ScS}, $\widetilde{S}^c\approx -y_HH_uH_d/\mu_S$ contributes
to $\Delta V(H)$ at one-loop level.
Note that the first two lines in \eq{effPot} are nothing but the
MSSM Higgs potential, while the two terms in the third line
correspond to the tree-level and one-loop corrections induced by
the heavy fields $\{\widetilde{S},\widetilde{S}^c\}$ and
$\{N,N^c\}$. The quartic term ``$y_H^2|H_uH_d|^2$'' in \eq{hsPot}
is canceled out, and so as seen from \eq{effPot}, the tree-level
corrections remain quite suppressed by heavy mass parameters.
As will be seen later, however, the one-loop
correction $\Delta V(H)$ can be relatively large since it originates from other sector rather than the MSSM. From now on, we will focus on the radiative correction, even if the tree-level quartic terms might be helpful for raising the Higgs mass in other parameter space violating Eq.~(\ref{hierarchy}).

The one-loop correction $\Delta V(H)$ in \eq{effPot} is just given
by \eq{1-loopPot}, but the $M_N$ in its expression should be
replaced by \dis{ \label{M_N} M_N^2\approx
\mu_N^2-\left(y_Hy_N\frac{\mu_N}{\mu_S}\right)h_uh_d 
+\frac{y_H^2y_N^2}{4\mu_S^2}(h_uh_d)^2
~, }
using \eq{ScS}. Here $h_{u,d}$ is the real component of $H_{u,d}$,
Re$H_{u,d}\equiv\frac{1}{\sqrt{2}}h_{u,d}$. 
We ignored the imaginary components of them.  
Thus, the expression
of $\Delta V(H)$ here is exactly the same as that of
Ref.~\cite{KP}. In Ref.~\cite{KP}, $\{N,N^c\}$ are integrated out
after $\{S,S^c\}$. As pointed out in Ref.~\cite{KP}, however, the
result should be insensitive to the sequence of the decouplings,
since the mass scales of $\{N,N^c\}$ and $\{S,S^c\}$ are not much
hierarchical.

We note that the similarity between the one-loop effective
potential of Eq.(\ref{1-loopPot}) with \eq{M_N} and that of the
footnote 1 in Introduction, which is the radiative Higgs potential
for a simple case of extra vectorlike matter. Accordingly, one can
expect that the Higgs mass is raised in our case through a similar
way to the case of extra vectorlike matter. The most important
difference between these two scenarios is that the fields
circulating along the loops are MSSM singlets in our case, while they
are charged fields under the SM in the extra vectorlike matter
case. In our case, lower scale mass parameters can be taken for
e.g. alleviating the tuning problem, but the LHC constraint on the
extra colored particles can be avoided unlike the extra vectorlike
matter case.

%
\begin{figure}
\begin{center}
\includegraphics[width=0.8\linewidth]{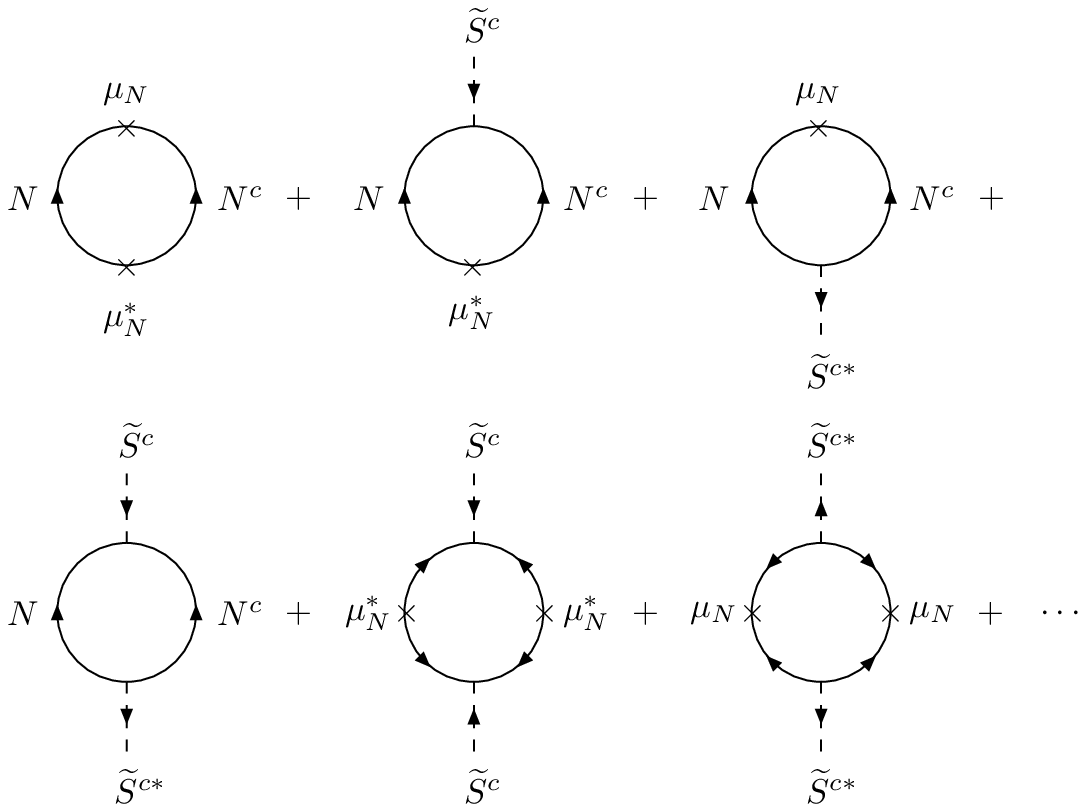}
\end{center}
\caption{Some contributions to the one-loop effective potential
of $\widetilde{S}^c$. Here we present only the diagrams of the
fermionic loops. By infinite summation of the amplitudes for all
the relevant one-loop diagrams, the Coleman-Weinberg's effective
potential for $\widetilde{S}^c$ can be obtained. Below the mass
scale of $\widetilde{S}^c$, the low energy effective Higgs
potential can be obtained by integrating out $\widetilde{S}^c$, in
which ``$H_uH_d$'' is attached to $\widetilde{S}^c$ in this setup.}
\end{figure}
%
%

In fact, the Coleman-Weinberg's one-loop effective potential,
$\Delta V(\widetilde{S}^c)$ of Eq.~(\ref{1-loopPot}) with
Eq.~(\ref{M_NSc}), can be obtained by taking infinite summation of
all possible one-loop diagrams, in which arbitrary numbers of
$\widetilde{S}^c$ are attached on the loop as the external legs
\cite{CW}. See the diagrams of FIG. 1, in which only the diagrams
of the fermionic loops are presented.\footnote{The second and third diagrams in FIG. 1 correspond to the tadpole of $\widetilde{S}^{c(*)}$, and the last two ones to $(\widetilde{S}^{c(*)})^2$ in the scalar potential. Although such terms are absent in \eq{hsPot}, they are radiatively induced. It is because \eq{superPot} might not be fully general in view of the symmetry. As mentioned in section \ref{sec2}, however, \eq{superPot} should be regarded as a low energy effective superpotential, and so its form is completely determined by a UV model embedding it. We will propose a UV model in section \ref{sec6}.  
}  
The diagrams with bosons in
the loops should be also considered.
In the effective operators
valid below the mass scale of $\widetilde{S}^c$, however,
$\widetilde{S}^c$ should appear as internal legs. As seen from
\eq{hsPot}, $\widetilde{S}^{c*}$ interacts only with $H_uH_d$ at
the tree-level, the external legs of the heavy field
$\widetilde{S}^c$ in FIG. 1 can couple to $H_uH_d$ 
at one-loop level. See FIG. 2-(b). Accordingly, $\Delta
V(\widetilde{S}^c)$ is converted to $\Delta V(H)$ below the mass
scale of $\widetilde{S}^c$. In fact, $\widetilde{S}^{c*}$ and
$\widetilde{S}$ are mixed and $\widetilde{S}$ is also coupled to
$H_uH_d$ via the $B_S$ and $A_S$ terms. Thus, $\widetilde{S}^c$
can couple to $H_uH_d$ through $\widetilde{S}$. However, this
possibility is more suppressed due to the hierarchical mass
relation in \eq{hierarchy}.

In this scenario, a nonzero radiative correction to the Higgs
mass squared is generated by the mass splitting of $\{N,N^c\}$ in
the hidden sector. The hidden sector in this model, thus, plays
the role of a mass generation sector of the Higgs. As seen in FIG. 2-(b), the nonzero mass effect is transmitted to the Higgs
through the messenger $\widetilde{S}^c$, which is actually a
mediator of the Higgs mass effect. The Higgs mass term generated in this
way can be meaningful only below the mass scale of
$\widetilde{S}^c$ ($\approx \mu_S$), because it can be regarded as
a local operator below the scale of $\mu_S$. Since $\widetilde{S}^c$ is a particle integrated out in the effective potential, its mass ($\approx \mu_S$) cannot be taken lighter than the mass of the Higgs, which is the particle of the external legs in the relevant diagrams, satisfying the classical equation of motion.

Figures 2-(a) and (b) show the typical diagrams for the radiatively
generated Higgs potentials by the top quarks in the MSSM and the
singlets in our case, respectively. They are compared to each
other. Actually, FIG. 2-(b) contributes to the renormalization of
$B_\mu$ term, while FIG. 2-(a) to the renormalization of the
$m_2^2$. The basic structures of the loops in the two diagrams are
the same. Roughly, the diagram of FIG. 2-(b) is estimated as
$(H_uH_d)(y_H\mu_S^*)\frac{1}{\mu_S^2}[n\times {\rm
Loop}]\mu_N^*$, while FIG. 2-(a) as $H_uy_t[3\times {\rm
Loop}]y_t^*H_u^*$, where the ``Loop'' means the common calculation
of the loops in the diagrams.

%
\begin{figure}
\begin{center}
\includegraphics[width=0.7\linewidth]{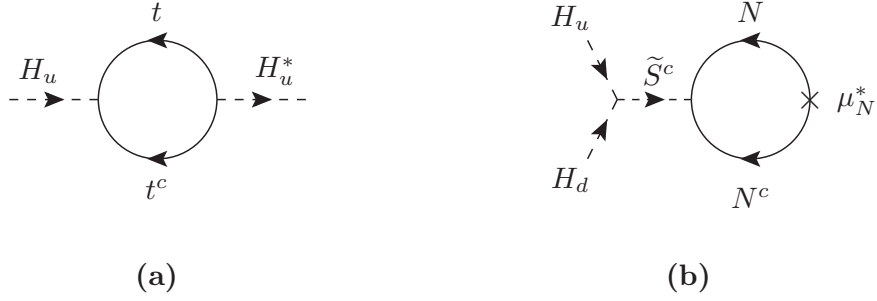}
\end{center}
\caption{{\bf (a)} A contribution to the radiatively induced
effective Higgs potential by the top quarks in the MSSM. {\bf (b)}
A contribution to the radiatively induced effective Higgs
potential by the singlets. It is compared with the diagram (a).
The basic structures of the loops in (a) and (b) are the same. The
trilinear scalar coupling in (b) comes from the cross term of
$|\partial{W}/\partial{S}|^2$. Radiatively generated mass in the
$\{N,N^c\}$ sector is transmitted to the Higgs through the
mediation by $\widetilde{S}^c$.}
\end{figure}
%
%

The radiative correction $\Delta V(H)$ given with
Eqs.~(\ref{1-loopPot}) and (\ref{M_N}) can be expanded in 
powers of $h_u$ and $h_d$ as follows: \dis{ \label{expansion}
\Delta V(H)=\Delta V^{\rm ren}_{(0,0)}+\Delta
V^{\rm ren}_{(1,1)}h_uh_d+\frac{1}{2!2!}\left\{\Delta V^{\rm ren}_{(2,2)}+\Delta V^{\rm phy}_{(2,2)}\right\}(h_uh_d)^2 +\cdots ,
} where the coefficients, $\Delta V^{\rm ren}_{(0,0)}$ [$\equiv
\Delta V(H)|_{h_u=h_d=0}$], $\Delta V^{\rm ren}_{(1,1)}$ [$\equiv
\partial_{h_u}\partial_{h_d}\Delta V(H)|_{h_u=h_d=0}$] and $\{\Delta
V^{\rm ren}_{(2,2)}+\Delta
V^{\rm phy}_{(2,2)}\}$ [$\equiv\partial^2_{h_u}\partial^2_{h_d}\Delta
V(H)|_{h_u=h_d=0}$] are estimated as
\begin{eqnarray} \label{coeff}
&&\Delta V^{\rm ren}_{(0,0)}= \frac{n}{16\pi^2}\bigg[\left(\mu_N^2+\widetilde{m}_{N}^2\right)^2\left\{{\rm
log}\left(\frac{\mu_N^2+\widetilde{m}_{N}^2}{\Lambda^2}\right)-\frac32\right\}
 -\mu_N^4\left\{{\rm
log}\left(\frac{\mu_N^2}{\Lambda^2}\right)-\frac32\right\} \bigg] ,
\nonumber \\
&&\Delta V^{\rm ren}_{(1,1)}= \frac{-n}{8\pi^2}\left(y_Hy_N\frac{\mu_N}{\mu_S}\right)\left[\left(\mu_N^2+\widetilde{m}_N^2\right)
\left\{{\rm log}\left(\frac{\mu_N^2+\widetilde{m}_N^2}{\Lambda^2}\right)-1\right\}-\mu_N^2\left\{{\rm log}\left(\frac{\mu_N^2}{\Lambda^2}\right)-1\right\}\right] ,
\nonumber \\
&&\Delta V^{\rm ren}_{(2,2)}= \frac{n}{8\pi^2}\left(\frac{y_Hy_N}{\mu_S}\right)^2\left[\left(\mu_N^2+\widetilde{m}_N^2\right)
\left\{{\rm log}\left(\frac{\mu_N^2+\widetilde{m}_N^2}{\Lambda^2}\right)-1\right\}-\mu_N^2\left\{{\rm log}\left(\frac{\mu_N^2}{\Lambda^2}\right)-1\right\}\right] ,
\nonumber \\
&&\Delta V^{\rm phy}_{(2,2)}= \frac{n}{4\pi^2}\left(y_Hy_N\frac{\mu_N}{\mu_S}\right)^2
{\rm log}\left(\frac{\mu_N^2+\widetilde{m}_N^2}{\mu_N^2}\right) .
\end{eqnarray}
Note that the coefficients of $h_u$, $h_d$, $h_u^2$, $h_d^2$, $h_u^3$, $h_d^3$,
$h_u^2h_d$, and $h_uh_d^2$ in \eq{expansion} are all zero, and the parts of
``$\cdots$'' are much suppressed by the higher powers of
$(y_Hy_Nh_{u,d}^2/\mu_S\mu_N)$. $\Delta V^{\rm ren}_{(0,0)}$ in
Eqs.~(\ref{expansion}) or (\ref{coeff}) just adds positive vacuum energy
as seen from the first diagram of FIG. 1, which is a result of
SUSY breaking.


As the (s-)top quark loops renormalize the soft mass squared of
the Higgs, $m_2^2$ in the MSSM, the diagram of
FIG. 2-(b) or
$\Delta V^{\rm ren}_{(1,1)}$ term in \eq{expansion} renormalizes the $B_\mu$
term in \eq{effPot} ($B_\mu\mu\equiv m_3^2$),
$m_3^2(\Lambda)=m_3^2-\Delta m_3^2$, where \dis{ \label{m3} \Delta
m_3^2\approx
\frac{n}{8\pi^2}\left(y_Hy_N\frac{\mu_N}{\mu_S}\right)\left[\left(\mu_N^2+\widetilde{m}_N^2\right)
\left\{{\rm
log}\left(\frac{\mu_N^2+\widetilde{m}_N^2}{\Lambda^2}\right)-1\right\}-\mu_N^2\left\{{\rm
log}\left(\frac{\mu_N^2}{\Lambda^2}\right)-1\right\}\right] . }
Since $\widetilde{S}^c$ plays the role of the messenger  relating $\{H_u,H_d\}$ and $\{N,N^c\}$, 
the mass scale of $\widetilde{S}^c$ ($\approx\mu_S$) is the messenger scale for inducing $\Delta m_3^2$. Below the $\mu_S$ scale, thus, FIG. 2-(b) can effectively be a irreducible diagram, and 
$\Delta m_3^2$ in \eq{m3} can be regarded as a local operator.  
Hence, $\Delta m_3^2$ in \eq{m3} is valid below $\mu_S$, in which 
$\widetilde{S}^c$ as well as $\{N,N^c\}$ are decoupled.
Thus, we set $\Lambda=\mu_S$ at lower energies. 

With the correction \eq{m3}, one of the tree-level extremum
conditions in the Higgs potential is modified as\footnote{The
extremum conditions in the MSSM are $m_1^2+|\mu|^2=m_3^2{\rm
tan}\beta-\frac{M_Z^2}{2}{\rm cos2\beta}$ and
$m_2^2+|\mu|^2=m_3^2{\rm cot}\beta+\frac{M_Z^2}{2}{\rm cos2\beta}$
at the tree-level, which can be recast into
$-2m_3^2=(m_1^2-m_2^2){\rm tan}2\beta +M_Z^2{\rm sin}2\beta$ and
$|\mu|^2=(m_2^2{\rm sin}^2\beta-m_1^2{\rm cos}^2\beta)/({\rm
cos}2\beta)-\frac12M_Z^2$ \cite{MSSM}.}
\begin{eqnarray} \label{extrm}
-2m_3^2=(m_1^2-m_2^2)~{\rm tan}2\beta + M_Z^2~{\rm sin}2\beta -2\Delta m_3^2 .
\end{eqnarray}
In order to avoid a fine-tuning among the parameters, $2\Delta
m_3^2$ needs to be comparable with other terms in \eq{extrm}, when
the parameters are chosen to explain the Higgs mass of 125 GeV. If
$\Delta m_3^2$ is too large, it should be properly canceled by 
other terms, being equated with  $M_Z^2{\rm sin}2\beta$ in
\eq{extrm}. Then, the tuning is roughly estimated by the
hierarchy, $M_Z^2~{\rm sin}2\beta/(2\Delta m_3^2)$.


The $\Delta V^{\rm ren}_{(2,2)}$ term in \eq{expansion}, 
which renormalizes the $(\widetilde{m}^2_{S^c}/\mu_S^2)y_H^2|H_uH_d|^2$ term in \eq{effPot}, originates from a quadratic term included in $\Delta V(\widetilde{S}^c)$ in \eq{1-loopPot}, 
\dis{ \label{renMSC}
\frac{ny_N^2}{8\pi^2}
\left[
\left(\mu_N^2+\widetilde{m}_N^2\right)
\left\{{\rm log}\left(\frac{\mu_N^2+\widetilde{m}_N^2}{\Lambda^2}\right)-1\right\}-\mu_N^2\left\{{\rm log}\left(\frac{\mu_N^2}{\Lambda^2}\right)-1\right\}\right]
|\widetilde{S}^c|^2 ,
}
which contributes to renormalization of the tree-level soft mass term, $\widetilde{m}_{S^c}^2(\Lambda)|\widetilde{S}^c|^2$ in \eq{hsPot}. 
Below the $\mu_S$ scale, $|\widetilde{S}^c|^2$ in \eq{renMSC} can be replaced by $(y_H/2\mu_S)^2(h_uh_d)^2$ as discussed before.
The structure of \eq{renMSC} should be exactly the same as the radiative correction of $m_2^2$ in the MSSM Higgs sector by the (s-)tops loops, as seen from the similarity of the fourth diagram in FIG. 1 and FIG. 2-(a). 
The mass term of $\widetilde{S}^c$ in the scalar potential \eq{hsPot}
is given by the summation of the above quadratic term \eq{renMSC} [$\equiv \delta\widetilde{m}^2_{S^c}(\Lambda)|\widetilde{S}^c|^2$], which comes from $\Delta V(\widetilde{S}^c)$ in \eq{1-loopPot}, and the tree-level soft mass term, which is also renormalization scale dependent. 
Inserting the RG solution of $\widetilde{m}^2_{S^c}$ in the tree-level soft mass squared $\widetilde{m}^2_{S^c}(\Lambda)$,
$\widetilde{m}^2_{S^c}(\Lambda)+\delta \widetilde{m}^2_{S^c}(\Lambda)$ yields the low energy  ($\Lambda<\mu_N$) value  of the renormalized $\tilde{m}^2_{S^c}$ 
in its RG evolution \cite{CQW}. As discussed already above \eq{hierarchy}, it was assumed to be relatively quite smaller than $\mu_S^2$ in \eq{hierarchy}: 
\dis{
\widetilde{m}^2_{S^c}(\Lambda=\mu_S)+\delta\widetilde{m}^2_{S^c}(\Lambda=\mu_S) 
\ll \mu_S^2 . 
} 
Note that the bosonic and fermionic modes of $\{N,N^c\}$ are all decoupled below the $\mu_N$ scale, and so $\widetilde{m}^2_{S^c}(\Lambda)+\delta\widetilde{m}^2_{S^c}(\Lambda)$ becomes frozen below $\mu_N$. 
Therefore, the $\Delta V^{\rm ren}_{(2,2)}$ term of \eq{coeff} in the scalar potential ensures the smallness of the tree-level quartic term in \eq{effPot} at the $\mu_S$ scale.\footnote{If the hierarchy Eq.~(\ref{hierarchy}) is violated, the tree-level $|H_uH_d|^2$ term can be helpful for raising the Higgs mass, but its effect is smaller than that of the NMSSM. Thus, a large radiative correction by large Yukawa couplings introducing a new source like $\{N,N^c\}$ is still needed.}


By comparing the quartic term, $\frac{1}{2!2!}\Delta V^{\rm phy}_{(2,2)}(h_uh_d)^2$ in \eq{expansion} with the scalar
potential in the NMSSM, $V\supset
\lambda^2|H_uH_d|^2=\frac{\lambda^2}{4}(h_uh_d)^2$, one can see
that $\Delta V^{\rm phy}_{(2,2)}$ in \eq{coeff} plays the role of
$\lambda^2$ of the NMSSM. Since we saw that the Higgs mass
correction to the lightest Higgs mass in the NMSSM is given by
$\lambda^2\times (v_H{\rm sin}2\beta)^2$ in \eq{HmassNMSSM}, we
can readily get the radiative correction $\Delta m_h^2$ in our
case:
\dis{ \label{mh} \Delta m_h^2\approx
\frac{n}{4\pi^2}\left(y_Hy_N\frac{\mu_N}{\mu_S}\right)^2\left(v_H^2{\rm
sin}^22\beta\right){\rm
log}\left(\frac{\mu_N^2+\widetilde{m}_N^2}{\mu_N^2}\right) . }
Note that $\mu_S$ originates from the propagator of
$\widetilde{S}^c$ in the diagram, while $\mu_N$ from the mass
insertion. Thus, the mass term correction by $\Delta m_h^2$ can be also 
a local operator below the messenger scale $\mu_S$. Since
the mass squared of $\widetilde{S}^c$ [$> (300~{\rm GeV})^2$] is
much heavier than the Higgs mass squared, $\Delta m_h^2$ in
\eq{mh} indeed can be the Higgs mass correction at
low energies.
For discussion of the consistency of the model
above the $\mu_N$ energy scale, one should return to
%
\eq{superPot},
in which $y_N$ can be of order unity.
%
By including \eq{mh}, thus, the CP even lightest Higgs
mass squared is modified as
\dis{ \label{fullmh} m_h^2\approx
M_Z^2{\rm cos}^22\beta
+\left\{\frac{\widetilde{m}_{S^c}^2}{\mu_S^2}-\frac{|A_S-B_S|^2}{\widetilde{m}_S^{2}+\mu_S^2}\right\}(y_H^2v_H^2{\rm
sin}^22\beta) + \Delta m_h^2|_{\rm MSSM} + \Delta m_h^2 . }
Due to 
the hierarchy \eq{hierarchy}, the classical correction is suppressed.  

As shown in Ref.~\cite{KP}, the Higgs mass of 125 GeV can be
explained with Eqs.~(\ref{fullmh}) or (\ref{mh}) in the parameter
space, \dis{ 0.2\lesssim y_H\lesssim 0.7 \quad {\rm or}\quad
3\lesssim {\rm tan}\beta\lesssim 10 , } without the mixing effect,
if the soft mass of the s-top is around 500 GeV [or $\Delta
m_h^2|_{\rm MSSM}\approx (66~{\rm GeV})^2$]. Thus, even
$0.2\lesssim y_H\lesssim 0.5$ or $3\lesssim {\rm tan}\beta\lesssim
10$, which is the excluded region in the NMSSM, can still be
consistent with the 125 GeV Higgs mass, when the radiative
correction of the Higgs mass is supported by the MSSM singlet
fields.

For the typical three classes,
$\mu_S\lesssim\widetilde{m}_N\lesssim\mu_N$ (Case A),
$\mu_S\lesssim\mu_N\lesssim\widetilde{m}_N$ (Case B), and
$\mu_S\lesssim\widetilde{m}_N\approx\mu_N$ (Case C), the radiative
corrections in Eqs.~(\ref{mh}) and (\ref{m3}) are approximated as
follows:
\noindent
\begin{eqnarray}
\label{caseA}
\left\{
\begin{array}{c}
\Delta m_h^2\approx \frac{n}{4\pi^2}\left(v_H^2{\rm sin}^22\beta\right)\left[\left(y_Hy_N\frac{\mu_N}{\mu_S}\right)^2\frac{\widetilde{m}_N^2}{\mu_N^2}\right]
\\
{\Delta m_3^2}\approx \frac{n}{4\pi^2}{\widetilde{m}_N^2}\left[\left(y_Hy_N\frac{\mu_N}{\mu_S}\right)
{\rm log}\left(\frac{\mu_N}{\mu_S}\right)\right]
\end{array}
\right. \quad{\rm for}~~\widetilde{m}_N\lesssim\mu_N~~({\rm Case~A}),
\end{eqnarray}
\begin{eqnarray}
 \label{caseB}
\left\{
\begin{array}{c}
\Delta m_h^2\approx \frac{n}{4\pi^2}\left(v_H^2{\rm sin}^22\beta\right)\left[\left(y_Hy_N\frac{\mu_N}{\mu_S}\right)^2{\rm log}\left(\frac{\widetilde{m}_N^2}{\mu_N^2}\right)\right]
\\
{\Delta m_3^2}\approx \frac{n}{4\pi^2}{\widetilde{m}_N^2}\left[\left(y_Hy_N\frac{\mu_N}{\mu_S}\right)
{\rm log}\left(\frac{\widetilde{m}_N}{\mu_S}\right)\right]
\end{array}
\right. \quad{\rm for}~~\mu_N\lesssim\widetilde{m}_N~~({\rm Case~B}),
\end{eqnarray}
\begin{eqnarray}
 \label{caseC}
\left\{
\begin{array}{c}
\Delta m_h^2\approx \frac{n}{4\pi^2}\left(v_H^2{\rm sin}^22\beta\right)\left[\left(y_Hy_N\frac{\mu_N}{\mu_S}\right)^2{\rm log}2\right]
\\
{\Delta m_3^2}\approx \frac{n}{4\pi^2}{\widetilde{m}_N^2}\left[\left(y_Hy_N\frac{\mu_N}{\mu_S}\right)
\left\{{\rm log}\left(2\frac{\mu_N}{\mu_S}\right)-\frac12\right\}\right]
\end{array}
\right. \quad{\rm for}~~\mu_N\approx\widetilde{m}_N~~({\rm Case~C}) .
\end{eqnarray}
In order to avoid a serious fine-tuning among the soft parameters
in \eq{extrm}, $\Delta m_3^2/v_H^2$ should not be too much larger
than unity.
From the above equations, it roughly means
$\widetilde{m}_N\lesssim 2\pi v_H\approx 1$ TeV. Hence, $\widetilde{m}_N$
should be quite smaller than 1 TeV. In the next section,
we will discuss this issue in more detail.

\section{125 GeV Higgs Mass with the Least Tuning} \label{sec4}

In this section, we study the least tuning condition, under which
the tuning in the Higgs sector is minimized for a given $\Delta
m_h^2$. For simple presentations, we parametrize the radiative
corrections in Eqs.~(\ref{mh}) and (\ref{m3}) as follows: \dis{
\label{FG} &\qquad\qquad\qquad\qquad F^2\equiv\frac{\Delta
m_h^2}{f^2v_H^2}=R^2~{\rm log}\left(1+r^2\right) ,
\\
&G\equiv\frac{2\Delta
m_3^2}{g\mu_S^2}=R^3\left[\left(1+r^2\right)\left\{ {\rm
log}(1+r^2)+{\rm log}R^2-1\right\}- \left\{{\rm
log}R^2-1\right\}\right] , } where $R$, $r$, and $f^2$, $g$ are
defined as \dis{ \label{def} &~~~~~~ R\equiv \frac{\mu_N}{\mu_S}
~, \quad r\equiv \frac{\widetilde{m}_N}{\mu_N} ~, \quad {\rm and}
\\
&f^2\equiv \frac{n}{4\pi^2} y_H^2y_N^2 {\rm sin}^22\beta ~, \quad g\equiv \frac{n}{4\pi^2}y_Hy_N .
}

For the parameters chosen for the explanation of the Higgs mass
around 125 GeV, as mentioned above, a smaller $\Delta m_3^2$ is
more desirable to avoid a fine-tuning among the parameters in
\eq{m3}.
From now on, we will explore the conditions under which  $\Delta
m_3^2$ can be minimized for a given $\Delta m_h^2$ and other
parameters in the model. As seen from \eq{FG}, $R$ and $r$ are
related to each other for a given $F$. Accordingly, $G$ depends only
on $r$ or $R$ for a fixed $F$. Let us insert $F$ into $G$,
replacing $r$ by $R$ and $F$. For a given set of $\{\Delta m_h^2,
\mu_S^2, f^2, g\}$, thus, $G$ is recast as \dis{
G=R^3\left[e^{\frac{F^2}{R^2}}\left(\frac{F^2}{R^2}+{\rm
log}R^2-1\right)-\left({\rm log}R^2-1\right)\right] . }
Provided that $F$ is fixed, one can show that $G$ is minimized at
\dis{
R=\frac{F}{1+\epsilon_F}
}
where the small parameter $\epsilon_F$ is estimated as
\dis{
\epsilon_F\approx \frac{1-0.28 ~{\rm log}F^2}{8.87+4.31 ~{\rm log}F^2} .
}
$|\epsilon_F|$ is much smaller than unity in the most parameter range of $F$:
$|\epsilon_F|$ is smaller
than $0.3$ ($0.1$) for $0<|F|<0.16$ or $0.59<|F|$ ($0<|F|<1.9\times 10^{-3}$ or $1.08<|F|$).
%
From \eq{FG}, thus, $r$ and $G$ are determined when $G$ minimized:
\dis{ \label{minm}
&\qquad\qquad~~~ r^2\approx 1.72+5.44\epsilon_F ,
\\
&G\approx F^3\left[(1.72+0.28\epsilon_F)~{\rm
log}F^2+(1-\epsilon_F)\right] . }
For instance, $\epsilon_F\approx
0.05$, $R\approx 1.75$, $r\approx 1.35$,  and $G\approx 19.09$ for
$F=1.84$. From \eq{def}, it implies that
$\frac{\mu_N}{\mu_S}\approx 1.75$,
$\frac{\widetilde{m}_N}{\mu_N}\approx 1.35$, and $\Delta
m_3^2\approx (330~{\rm GeV})^2$ e.g. for $|\Delta m_h|= 91$ GeV,
$\mu_S=300$ GeV, $n=5$, $y_Hy_N=1$, and ${\rm sin}2\beta=0.8$.

Note that $0.3<r<1.8$ for $-0.3<\epsilon_F<0.3$ in \eq{minm}.  We
can see that $\mu_N$ and $\widetilde{m}_N$ need to be comparable
to each other in order to minimize $\Delta m_3^2$. However,
$\Delta m_3^2$ is {\it not} much sensitive to $\widetilde{m}_N/\mu_N$ ($=r$), only if $\widetilde{m}_N/\mu_N$ is
larger than unity, because $r$ logarithmically depends on the
constraint relation associated with $F$ in \eq{FG}.

In \eq{minm}, $G$ could be further minimized with a small $F$.
Since $\Delta m_h^2\approx m_h^2-M_Z^2{\rm cos}^22\beta-\Delta m_h^2|_{\rm MSSM}$, $F^2$ in \eq{FG} is minimized when ${\rm sin}^22\beta =1$ (or ${\rm tan}\beta=1$):
\dis{
F^2\approx\frac{m_h^2-M_Z^2-\Delta m_h^2|_{\rm MSSM}+M_Z^2~{\rm sin}^22\beta}
{\frac{n}{4\pi^2}(y_Hy_N)^2v_H^2~{\rm sin}^22\beta}
\geq \frac{m_h^2-\Delta m_h^2|_{\rm MSSM}}
{\frac{n}{4\pi^2}(y_Hy_N)^2v_H^2}~,
}
For $n=5$, $(y_Hy_N)=1$, and $\Delta m_h^2|_{\rm MSSM}=(66~{\rm
GeV})^2$ [which corresponds to $\widetilde{m}_t\approx 500$ GeV at
two-loop level],  thus, the minimum of $F^2$ is $(1.71)^2$, which
gives $G\approx 14.07$ or $\Delta m_3^2\approx (284~{\rm GeV})^2$.
To avoid another fine-tuning needed for minimizing the tuning,
however, we do not rigorously apply the least tuning condition.
Nonetheless, the tuning problem associated with the extremum
conditions can still be remarkably mitigated with relatively
smaller $\Delta m_3^2$, compared to the MSSM. For $\Delta m_3^2$
with other parameters, see FIG. 3 and 4.  Note that  even the soft
parameters much lighter than 500 GeV can explain the Higgs mass of
125 GeV.  Taking such light mass parameters is not in conflict with
the LHC experimental results unlike the extra matter scenario. It
is possible because the newly introduced particles are MSSM
singlets.


%
\begin{figure}
\begin{center}
\includegraphics[width=0.85\linewidth]{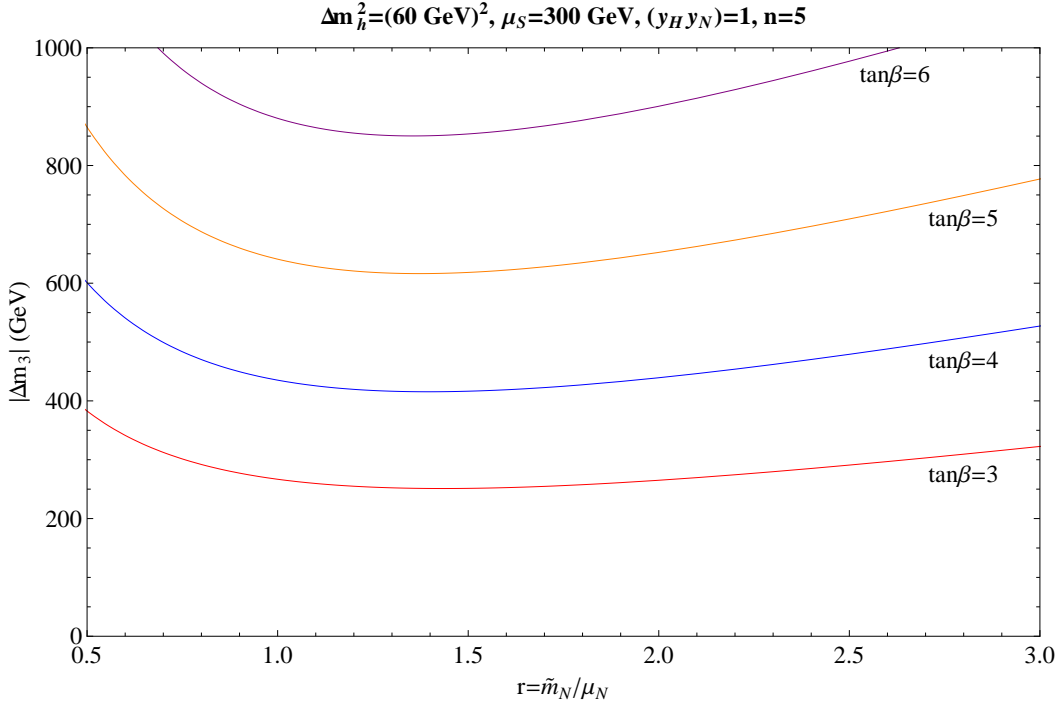}
\end{center}
\caption{ Radiative correction $|\Delta m_3|$ ($\equiv\sqrt{B_\mu\mu}$) vs. $\widetilde{m}_N/\mu_N$
for various values of ${\rm tan}\beta$. The radiative correction to the Higgs mass $\Delta m_h^2$ is set to  $(60~{\rm GeV})^2$. By $|\Delta m_h|_{\rm MSSM}\approx (68, 70, 75, 82)$ GeV for ${\rm tan}\beta=(6,5,4,3)$, thus, the Higgs mass get additional contributions from the (s-)top to be  125 GeV. 
They correspond to $\widetilde{m}_t\approx (530, 590, 780, 1300)$ GeV at two-loop level, when turning off the mixing effect of $({\tilde{t}_L,\tilde{t}_R})$ 
and the tree-level $|H_uH_d|^2$ terms in Eq.~(\ref{effPot}).
We fix the other parameters as shown in the figure.
} \label{Fig_mN1}
\end{figure}

\begin{figure}
\begin{center}
\includegraphics[width=0.85\linewidth]{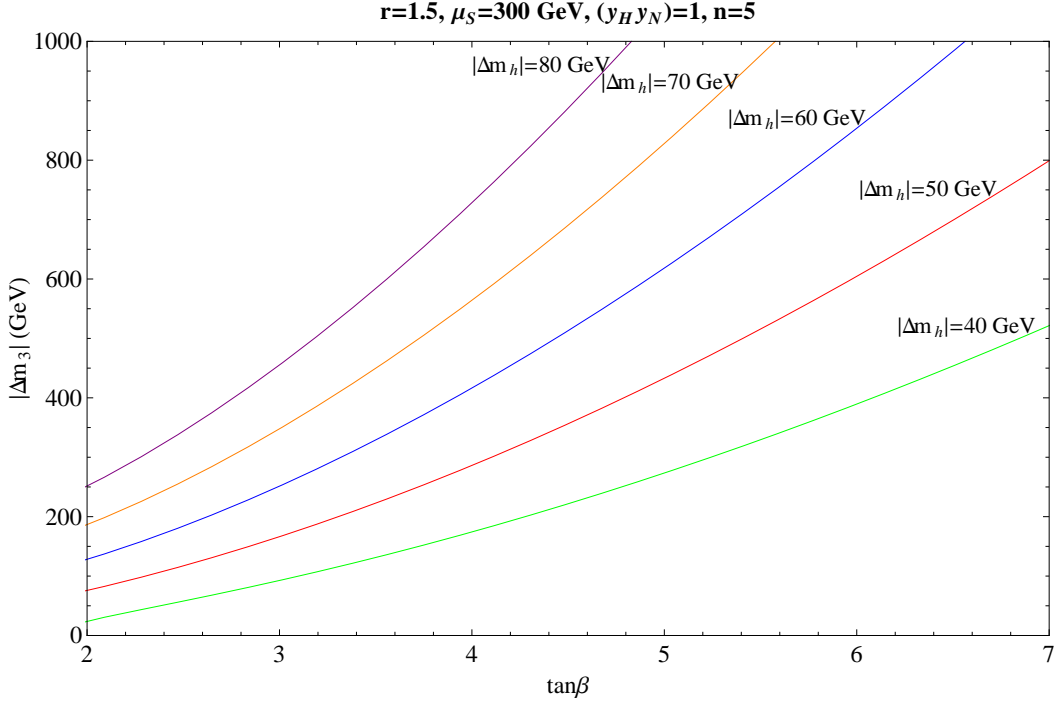}
\end{center}
\caption{ Radiative correction $|\Delta m_3|$ ($\equiv\sqrt{B_\mu\mu}$) vs. ${\rm tan}\beta$
for various values of $\Delta m_h^2$ around the least tuning points ($\widetilde{m}_N/\mu_N \approx1.5$).
$|\Delta m_h|=(80, 70, 60, 50, 40)$ GeV for ${\rm tan}\beta=5$ require the supplements of
$|\Delta m_h|_{\rm MSSM}\approx (47, 61, 70, 78, 83)$ GeV, respectively, by the (s-)top's
contributions.  They correspond to  $\widetilde{m}_t\approx (230, 390, 590, 940, 1400)$ GeV
at two-loop level, when turning off the mixing effect of $({\tilde{t}_L,\tilde{t}_R})$
and the tree-level $|H_uH_d|^2$ terms in Eq.~(\ref{effPot}). 
The other parameters are fixed as shown in the figure.
} \label{Fig_TanBeta1}
\end{figure}


Let us present the estimates of typical values of $\Delta m_3^2$
for the three classes defined in section \ref{sec3}, when $\Delta m_h^2$
and other parameters are given. In Case A, namely, for
$\widetilde{m}_N\lesssim \mu_N$, we have
\begin{eqnarray}
&&\qquad\qquad\qquad
\Delta m_3^2\approx \mu_S^2 \left[\frac{|\Delta
m_h|}{v_H}\right]^3\frac{g}{f^3r}~{\rm log}\left(\frac{|\Delta
m_h|}{fv_Hr}\right)
\\
&&\approx
(592~{\rm GeV})^2 \left[\frac{\mu_S}{300~{\rm GeV}}\right]^2
\left[\frac{|\Delta m_h|}{90~{\rm GeV}}\right]^3\left[\frac{1.14}{\sqrt{n}(y_Hy_N)^2{\rm sin}^32\beta}\right]\left[\frac{\frac{1}{r}~{\rm log}\left(\frac{|\Delta m_h|}{f v_Hr}\right)}{3~{\rm log}\left(\frac{3\cdot 90}{0.28\cdot 174}\right)}\right] .
\nonumber
\end{eqnarray}
In Case B, i.e. for $\mu_N\lesssim\widetilde{m}_N$,
\begin{eqnarray}
&&\qquad\qquad\qquad \Delta m_3^2\approx \mu_S^2 \left[\frac{|\Delta m_h|}{v_H}\right]^3\frac{g^{}r^2}{f^3({\rm log}r^2)^{3/2}}~{\rm log}\left(\frac{r|\Delta m_h|}{\sqrt{{\rm log}r^2}fv_H}\right)
\\ \nonumber
&&\approx (499~{\rm GeV})^2 \left[\frac{\mu_S}{300~{\rm GeV}}\right]^2
\left[\frac{|\Delta m_h|}{90~{\rm GeV}}\right]^3\left[\frac{1.14}{\sqrt{n}(y_Hy_N){\rm sin}^32\beta}\right]\left[\frac{\frac{r^2}{({\rm log}r^2)^{3/2}}~{\rm log}\left(\frac{r|\Delta m_h|}{\sqrt{{\rm log}r^2}fv_H}\right)}{2.76~{\rm log}\left(\frac{3\cdot 90}{1.48\cdot 0.28\cdot 174}\right)}\right] .
\end{eqnarray}
In Case C, i.e. for $\mu_N\approx\widetilde{m}_N$,
\begin{eqnarray}
&&\qquad\quad\quad
\Delta m_3^2\approx
\mu_S^2\left[\frac{|\Delta m_h|}{v_H}\right]^3\frac{g}{f^3({\rm log}2)^{3/2}}\left[{\rm log}\left(\frac{2|\Delta m_h|}{fv_H\sqrt{{\rm log}2}}\right)-\frac12\right]
\\
&&\approx
(342~{\rm GeV})^2\left[\frac{\mu_S}{300~{\rm GeV}}\right]^2
\left[\frac{|\Delta m_h|}{90~{\rm GeV}}\right]^3
\left[\frac{1.14}{\sqrt{n}(y_Hy_N)^2{\rm sin}^32\beta}\right]\left[\frac{{\rm log}\left(\frac{2|\Delta m_h|}{fv_H\sqrt{{\rm log}2}}\right)-\frac12}{0.99}\right] .
\nonumber
\end{eqnarray}

\section{Diphoton Decay Enhancement} \label{sec5}

According to the reports by the CMS and ATLAS \cite{CMS,ATLAS},
they both have observed an excess in the Higgs production and
decay to the diphoton channel, which is about 1.5 -- 2 times
larger than the SM expectation. On the other hand, the $ZZ$ and
$WW$ channels are quite compatible with the SM: \dis{ &
\frac{\sigma(gg\rightarrow h)\times {\rm
Br}(h\rightarrow\gamma\gamma)}{[\sigma(gg\rightarrow h)\times {\rm
Br}(h\rightarrow\gamma\gamma)]_{\rm SM}}\sim 1.5 - 2 ,
\\
&~\quad \frac{\sigma(gg\rightarrow h)\times {\rm Br}(h\rightarrow
VV)}{[\sigma(gg\rightarrow h)\times {\rm Br}(h\rightarrow
VV)]_{\rm SM}}\sim 1 , } where $V$ indicates $Z$ or $W$. In fact,
the excess at 8 TeV of the LHC slightly decreases compared to that
for 7 TeV. However, if the large excess in the diphoton decay
channel persists even after further more precise analyses with
more data, one must seriously consider the possibility of the
presence of new charged particles at low energies \cite{C,A}.

So far, we have regarded $\{N, N^c\}$ as vectorlike
$n$-dimensional representations of a hidden gauge group. In this
section, however, by slightly modifying the model, namely,
assigning additional  electromagnetic (or hyper) charges, $Q_N$
and $-Q_N$, respectively to $N$ and $N^c$, we attempt to explain
the excess of the diphoton decay rate of the Higgs under the
assumption that the enhanced diphoton decay rate of the Higgs will
survive. Thus, the mechanism of the Higgs mass enhancement and
mitigating the fine-tuning can be closely  associated with the
excess of the diphoton decay rate of the Higgs in our framework.
Since $\{N, N^c\}$ do not carry any SU(3)$_c$ and SU(2)$_L$
quantum numbers, they would not affect the Higgs production rate  at the LHC
$gg\rightarrow h$, and decay rate $h\rightarrow WW$. Also they do not much perturb the tree-level decay rate of
$h\rightarrow ZZ$. With $\{N, N^c\}$ carrying U(1)$_Y$
charges, however, the gauge coupling unification in the MSSM is
spoiled, unless an exotic normalization of U(1)$_Y$ is supported
in a UV theory. It is the cost for the explanation of the diphoton
excess of the Higgs.

As discussed in Ref.~\cite{C}, e.g. by extra vectorlike charged
leptons, the sizable enhancement of $h\rightarrow\gamma\gamma$ can
be successfully achieved, if the coefficient of the dimension five
interaction between the Higgs boson and the fermion,
$(c_f/\Lambda)H^\dagger H\bar{f}f$ is negative. We can obtain a
similar operator by integrating out $\widetilde{S}^c$ in our framework:
\dis{ \label{L}
-{\cal L}_{\rm eff}= -\frac{y_Hy_N}{\mu_S}H_uH_dNN^c ~~ + ~~ {\rm h.c.} .
}
Here $N$, $N^c$ are the fermionic modes of the superfields
$\{N,N^c\}$ (Weyl fermions). They form a Dirac fermion, $f=(N,N^{c*})^T$.
Thus, $\{N,N^c\}$ get an additional mass coming from the
Higgs' vacuum expectation values (VEVs) apart from the bare mass $\mu_N$:
\dis{ \label{M}
M_N\approx \mu_N-y_Hy_N\frac{v_uv_d}{\mu_S} ,
}
where $v_{u,d}\equiv\langle H_{u,d}\rangle$.
It can also be obtained from \eq{M_NSc} and the solution of
$\widetilde{S}^c$ in \eq{ScS}.
Connecting the $N$, $N^c$ lines in \eq{L}, the operator
associated with the diagram in FIG. 2-(b) is reproduced.
The relevant diagram for $h\rightarrow\gamma\gamma$ is
obtained by attaching two photons to the loops.
Of course, the bosonic modes of $\{N,N^c\}$ also make a
contribution to $h\rightarrow\gamma\gamma$.
However, they less affect the decay, since they are relatively
heavier than the fermionic partners.
%
%
With Eqs.~(\ref{L}) and (\ref{M}), the enhancement factor
over the SM diphoton width \cite{C} is estimated in the
heavy Higgs decoupling limit as follows:
\dis{ \label{EF}
R_{\gamma\gamma}\approx \left|1-\frac{y_Hy_N}{\sqrt{2}}\frac{v_H^2{\rm sin}2\beta}{\mu_SM_N}\frac{nQ^2_N\left\{A_{1/2}(x_N)+{\cal O}\left(\frac{m_h}{\widetilde{m}_{N}}\right)\right\}}{A_1(x_W)+3\left(\frac{2}{3}\right)^2A_{1/2}(x_t)}\right|^2 ,
}
where $x_i\equiv 4m_i^2/m_h^2$.
$n$ denotes the dimension of
the representation of $\{N,N^c\}$ under a hidden gauge
group, 
and $(-)Q_N$ means the electromagnetic charge $N$ ($N^c$) carries.
Below the $WW$ threshold, the loop functions for the
vector boson  ($A_1$) and the fermion ($A_{1/2}$) are given by
\dis{
&A_1(x)=-x^2\left[2x^{-2}+3x^{-1}
+3\left(2x^{-1}-1\right)f(x^{-1})\right]  ,
\\
&\qquad A_{1/2}(x)=2x^2\left[x^{-1}
+\left(x^{-1}-1\right)f(x^{-1})\right] ,
}
where $f(x^{-1})\equiv {\rm arcsin}^2 x^{-1/2}$.
We ignore the contributions by the bosonic partners in $\{N,N^c\}$, which is just of order ${\cal O}\left(m_h/\widetilde{m}_{N}\right)$, because of their relatively heavier masses.

The main SM contributions, $A_{1}(x_W)$ by the $W$ boson and
$3(\frac{2}{3})^2A_{1/2}(x_t)$ by the top quark, which appear in
the denominator of \eq{EF} are $-8.32$ and $1.84$, respectively.
For constructive interference, thus, the sign of
$(y_Hy_N/\mu_SM_N)$ should be positive. See FIG. 5, in which we
display the contour plots for the enhancement factor over the SM
diphoton width in the $M_N$--$Q_N$ plane for the $n=5$ and $n=1$
cases.

%
\begin{figure}
\begin{center}
\includegraphics[width=0.48\linewidth]{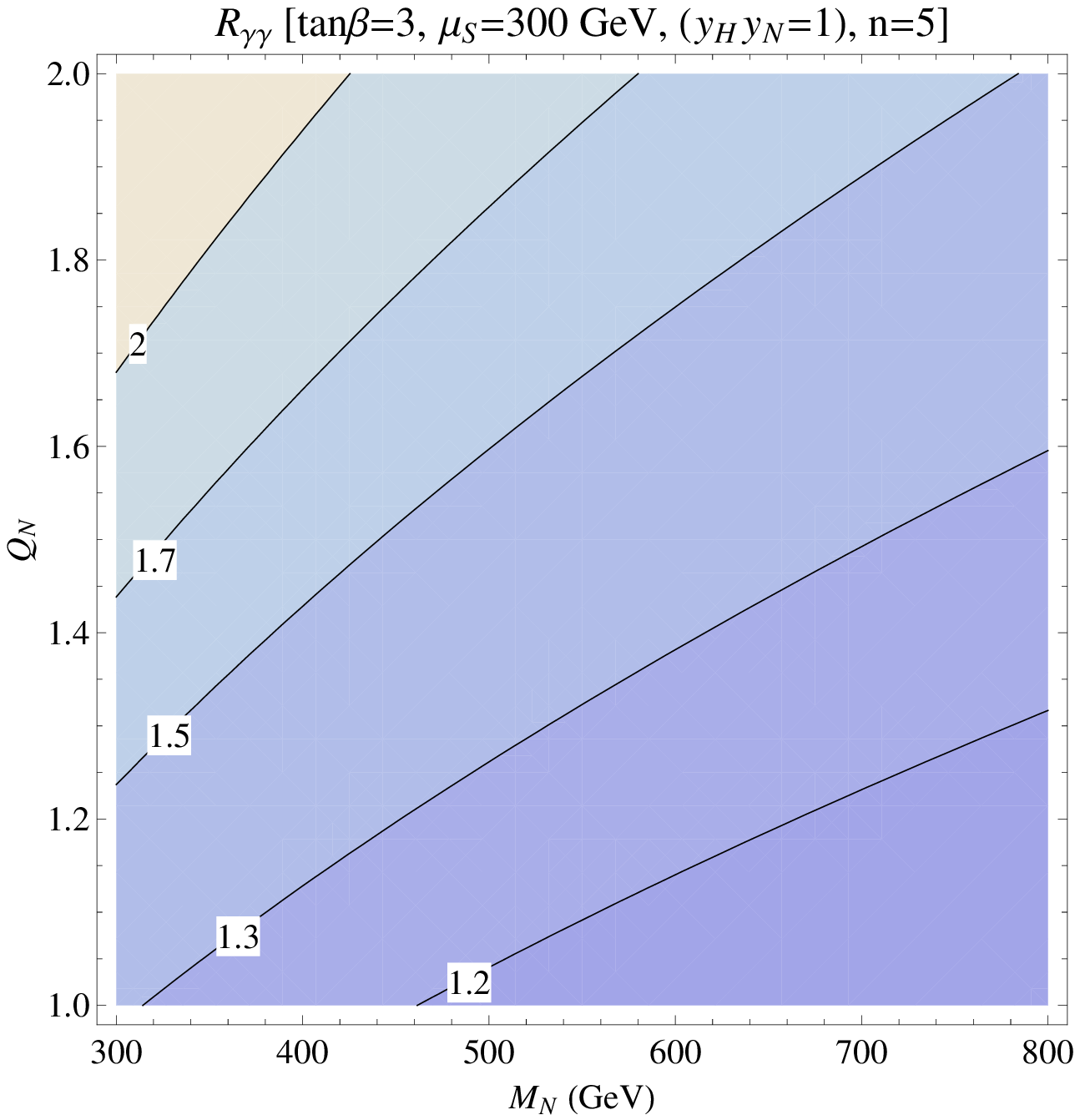}
\includegraphics[width=0.48\linewidth]{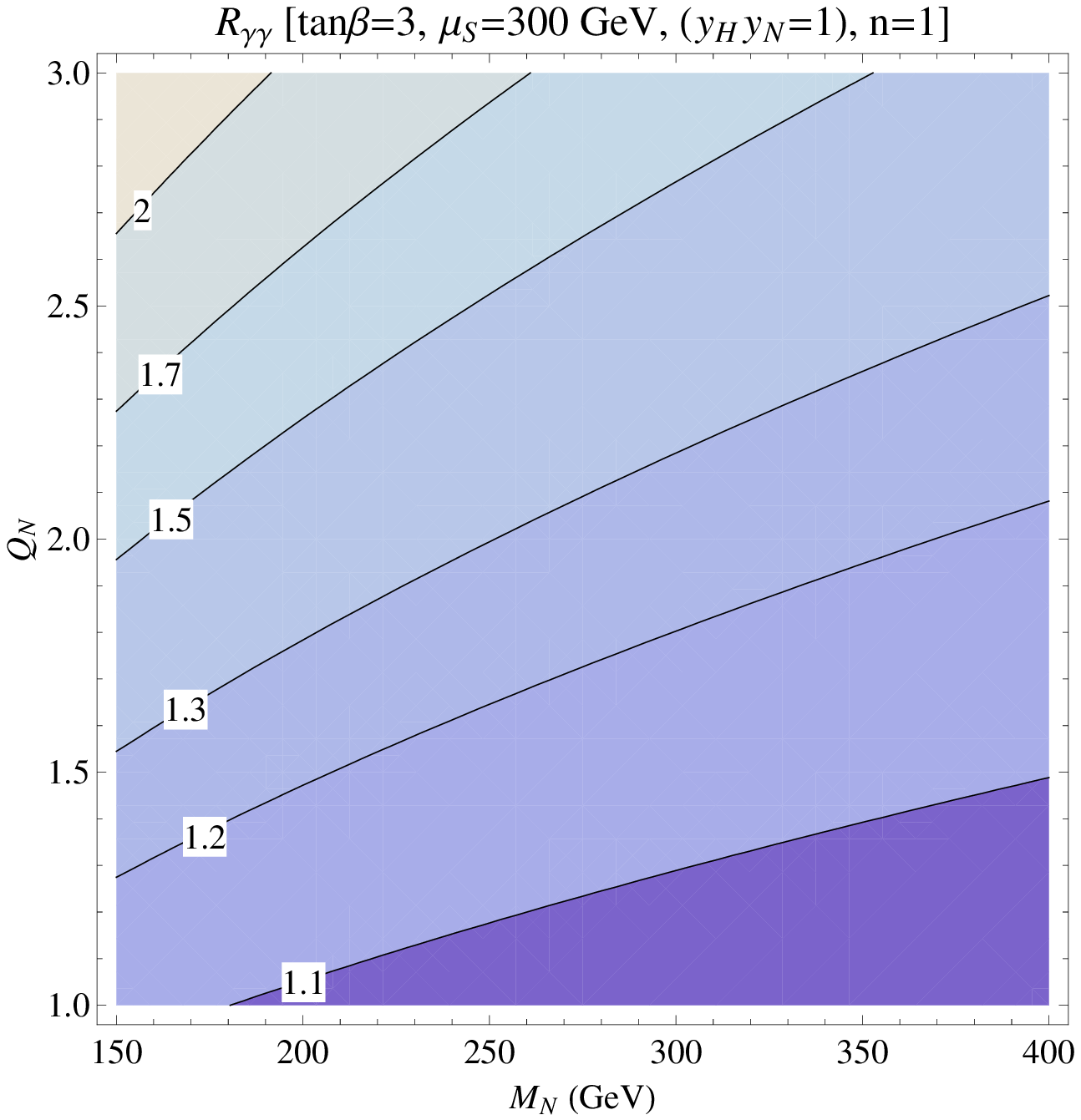}
\end{center}
\caption{Contour plots for the enhancement factor over the SM diphoton width in the $M_N$--$Q_N$ plane. We fix the other parameters as shown in each figure.} \label{Fig_Rrr}
\end{figure}
%
%

Only with \eq{superPot}, a considerable amount of $\{N,N^c\}$
would remain as cosmological relic, unless the reheating
temperature is very low, which is a disaster when they carry
electromagnetic charges. To avoid it, we discuss two possibilities
here. One could consider the possibility that $N$, $N^c$ condense
by the strong hidden gauge interaction as the quarks in QCD. Then,
only the neutral hadron would remain in our case, and it can decay
to the two photons as the pion $\pi^0$ in QCD. In this case, $M_N$
in \eq{EF} should be replaced by $8\pi^2f_N/3$, where $f_N$ is the
decay constant determined by confining of the hidden gauge
interaction. Alternatively, if $n=1$ and $Q_N=-2$, the
superpotential allows the interactions with the MSSM charged
lepton singlets, $N(e^c)^2$. Then, $N$, $N^c$, and
$\widetilde{N}$, $\widetilde{N}^c$ can decay eventually to $e^\pm$
and the neutralinos before nucleosynthesis starts even without the
assumption of the hidden confining gauge interaction.


\section{The Model} \label{sec6}

As mentioned in section \ref{sec2}, the superpotential \eq{superPot} should be embedded in the superpotential of a UV model, which permits more global symmetries. 
The singlets $S$ and $S^c$ should be charged under the global symmetries to avoid the tadpole problem associated with pure singlets \cite{tadpole1,tadpole2}. 
The global symmetries should be broken such that there is no remaining PQ symmetry at low energies, explaining the desired sizes of $\mu$, $\mu_S$, and $\mu_N$ in \eq{superPot}. 
If the PQ symmetry is broken at the scale of $\sqrt{m_{3/2}M_P}$ ($\sim 10^{10}$ GeV), the tadpole problem could be avoided \cite{tadpole1,JSY}. 

The effective superpotential \eq{superPot} can be deduced e.g.
from the following UV K${\rm\ddot{a}}$hler potential and the  superpotential:
\dis{ \label{UV}
&\qquad\quad ~~ K_{\rm UV}\supset \kappa\frac{X^\dagger}{M_P}SS^c
~~+~~{\rm h.c.} \,,
\\
&\qquad\quad W_{\rm UV}=y_HSH_uH_d+y_NS^cNN^c
\\
&+\frac{\lambda_1}{M_P}\Sigma_1^2H_uH_d+\frac{\lambda_2}{M_P}\Sigma_2^2NN^c
+\frac{\lambda_3}{M_P}\Sigma_1\Sigma_2\overline{\Sigma}^2 ,
\\
}
where $y_{H}$, $y_N$, $\kappa$ and $\lambda_i$ ($i=1,2,3$) are
dimensionless couplings, and $M_P$ denotes the reduced Planck mass
($=2.4\times 10^{18}$ GeV). 
The K${\rm\ddot{a}}$hler potential and the superpotential \eq{UV} respect the global symmetry, U(1)$_{\rm R}\times$U(1)$_{\rm PQ}$.
The global charges for the superfields are displayed in TABLE I.

\begin{table}[!h]
\begin{center}
\begin{tabular}
{c|cccccc|cccc}
{\rm Superfields}  &   $H_u$   &
 $H_d$  &  ~$N$  & ~$N^c$ & ~$S$  &
 ~$S^c$  & ~$\Sigma_1$  & ~$\Sigma_2$ & ~$\overline{\Sigma}$ & ~$X$   \\
\hline
U(1)$_{\rm R}$ & ~$0$ & ~$0$ & ~$0$ & ~$0$
 & ~$2$ & ~$2$ & ~$1$  & ~$1$ & ~$0$& ~$4$ \\
U(1)$_{\rm PQ}$ & $-\frac14$ & $-\frac14$ & ~$\frac12$ & ~$\frac12$ & ~$\frac12$ & $-1$ & ~$\frac14$ & $-\frac12$ & ~$\frac18$& $-\frac12$
%
\end{tabular}
\end{center}\caption{R and Pecci-Quinn charges of the superfields. The MSSM
{\it matter} superfields carry unit R charges, and  also 
PQ charges of $1/8$. $N$ and $N^c$ are assumed to be proper
$n$-dimensional vectorlike representations of a hidden gauge
group, under which all the MSSM fields are neutral. 
}\label{tab:Qnumb}
\end{table}

The $F$-component of the superfield $X$ is assumed to develop a VEV of order $m_{3/2}M_P$, breaking SUSY. 
Thus, the $\mu_S$ term of order $m_{3/2}$ in \eq{superPot} can be generated from the K${\rm\ddot{a}}$hler potential \eq{UV} \cite{GM}.  
By the ``$A$-term'' corresponding to the $\lambda_{3}$ terms in
\eq{UV} and the soft mass terms in the scalar potential, the VEVs of $\Sigma_{1,2}$ and
$\overline{\Sigma}$ of order $\sqrt{m_{3/2}M_P}$ ($\sim
10^{10}$ GeV) are generated at the minimum \cite{422}.
From the $\lambda_{1,2}$ terms in \eq{UV}, thus, ``$\mu$'' in the
MSSM, and also $\mu_N$ in \eq{superPot}, which are also of order $m_{3/2}$ \cite{Kim-Nilles}, are generated. 

The global symmetries are broken by the SUSY breaking effects: 
by the VEV of the $F$-component of $X$, the U(1)$_{\rm R}$ symmetry is broken to $Z_2$, which is identified with the matter parity in the MSSM, and    
due to the VEVs of $\{\Sigma_{1,2}, \overline{\Sigma}\}$, 
U(1)$_{\rm PQ}$ are completely broken at the intermediate scale.
Note that a tadpole term of $S^c$ in the superpotential can be induced after the global symmetries are broken [$W\supset S^c(\langle\overline{\Sigma}\rangle)^8/M_P^6\sim S^cm_{3/2}^4/M_P^2$], but it is extremely suppressed. 
Since $\Sigma_{1,2}$ and $\overline{\Sigma}$ carry accidental $Z_2\times Z_2^\prime$ charges of $(1,0)$ and $(0,1)$, respectively, 
a domain wall problem would potentially arise.
Hence, we assume that the  discrete symmetries were already broken
before or during inflation such that domain walls were diluted
away. If the reheating temperature is lower than $10^9$ GeV, the
$Z_2\times Z_2^\prime$ breaking vacuum can still be the minimum of
the potential also after inflation \cite{422}.

Finally, let us discuss the tadpole problem \cite{tadpole1} 
in this case. 
The K${\rm\ddot{a}}$hler potential e.g. for the Higgs fields takes the following form: 
\dis{ \label{Ksinglet}
K_{\rm UV}\supset \sum_{i=u,d} H_iH_i^\dagger
+ \frac{\alpha_iH_iH_i^\dagger}{M_P}\left\{S\left(\frac{\Sigma_1^\dagger}{M_P}\right)^2
+ {\rm h.c.}
\right\} 
+ \frac{\beta_iH_iH_i^\dagger}{M_P}\left\{S^c\left(\frac{\Sigma_2^\dagger}{M_P}\right)^2
+ {\rm h.c.} 
\right\} ,  
}
which is consistent with the quantum numbers listed in TABLE I. 
Note that $(\Sigma_{1,2}^\dagger/M_P)^2$ are accompanied with $S^{(c)}$ in \eq{Ksinglet}, since $S$, $S^c$ carry the global charges. They effectively suppress the coefficients $\alpha_i$ and $\beta_i$ with $(\langle \widetilde{\Sigma}_{1,2}\rangle/M_P)^2\sim m_{3/2}/M_P$. 
When SUSY is broken in the hidden sector, thus, the scalar potential and kinetic terms in SUGRA with \eq{Ksinglet} are
recast into 
\dis{ \label{Vsinglet}
&V_{\rm vis.}=K^{ij*}W_iW_j^*+m_{3/2}^2K^{ij*}K_iK_{j*}+ {\cal O}\left(\frac{m_{3/2}^2}{M_P^2}\right)
\\
&\supset \sum_{i=u,d}|H_i|^2\frac{m_{3/2}^3}{M_P^2}\bigg[\alpha_i^\prime\left\{\widetilde{S} + \widetilde{S}^* \right\} 
+ \beta_i^\prime\left\{\widetilde{S}^c + \widetilde{S}^{c*}\right\}\bigg] ,
}
where superscripts and subscripts in the K${\rm\ddot{a}}$hler and superpotential denote differentiations with respect to the scalar fields in SUGRA, and 
\dis{ \label{Lsinglet}
&{\cal L}_{\rm kin.}=K_{ij*}\partial_\mu z^i\partial^\mu z^{j*}
 \supset \sum_{i=u,d} |\partial_\mu H_i|^2 
\frac{m_{3/2}}{M_P^2}\bigg[\alpha_i^\prime\left\{\widetilde{S} + \widetilde{S}^* \right\} 
+ \beta_i^\prime\left\{\widetilde{S}^c + \widetilde{S}^{c*}\right\}\bigg] . 
}
$|H_i|^2$ in \eq{Vsinglet} and $|\partial_\mu H_i|^2$ in \eq{Lsinglet} introduce quadratic divergences in the loop integrals, 
inducing the tadpole terms of $\widetilde{S}$ and $\widetilde{S}^c$ in the Lagrangian,
\dis{
\Lambda_{\rm cutoff}^2 ~\frac{m_{3/2}^3}{M_P^2}\left\{\widetilde{S}^{(c)}+\widetilde{S}^{(c)*}\right\} , 
}
where we dropped the numerical factors.\footnote{  
The tadpole of $\widetilde{S}^c$ is renormalized by the superpotential sector as seen from the second and third diagrams in FIG. 1, when SUSY is broken. The tadpole of $\widetilde{S}$ is also similarly renormalized by $\{H_u,H_d\}$.}
Even if $\Lambda_{\rm cutoff}^2=M_P$, thus, the tadpole coefficients are just of order $m_{3/2}$ or smaller.  
With the minimal K${\rm\ddot{a}}$hler potential, moreover, such divergences are known to be canceled out at the one-loop level  \cite{Jain}.  
Accordingly, the shifts of the VEVs by the tadpoles, $\langle \delta \widetilde{S}\rangle$ and $\langle \delta \widetilde{S}^c\rangle$ are quite suppressed in our case, and so the gauge hierarchy is not destabilized by them.  
In this paper, hence, we neglect their effects.   
Note that were it not for the global symmetries, 
$(\Sigma_{1,2}^\dagger/M_P)^2$ are absent in \eq{Ksinglet}.  
Without the factors, we had extremely huge tadpole terms, $\Lambda_{\rm cutoff}^2\frac{m_{3/2}^2}{M_P}\{\widetilde{S}^{(c)}+\widetilde{S}^{(c)*}\}$, which 
destabilizes the gauge hierarchy, since $S$ couples to $H_u$ and $H_d$ at the tree level in the superpotential \eq{superPot} \cite{tadpole1}.

\section{Conclusion} \label{sec7}

We proposed a new type of the singlet extension of the MSSM in
order to raise the Higgs mass to 125 GeV with the alleviation of
the tuning associated with the light Higgs mass. Apart from the
(s-)top quark's contribution, the Higgs mass is radiatively
generated in a hidden sector because of the mass splitting of
hidden sector fields, and such an effect is transmitted to the
Higgs sector through the mediation by the messenger field
$\widetilde{S}^c$. Since the Higgs mass is raised 
by the superpotential parameters, lifting the
Higgs mass is quite efficient as in the extra matter scenario.
Unlike the extra matter scenario, however, our model is free from
the constraint on extra colored particles with order-one Yukawa
couplings to the Higgs, which is associated with the production
and decay rates of the Higgs at the LHC \cite{4thfamily}.

As shown in our previous paper \cite{KP}, the parameter space for
125 GeV Higgs mass can be enlarged compared to the original form
of the NMSSM, and so even $0.2\lesssim y_H\lesssim 0.5$ or
$3\lesssim {\rm tan}\beta\lesssim 10$, which is excluded region in
the NMSSM, can explain the 125 GeV Higgs mass with a relatively
light s-top ($\sim 500$ GeV) but without considering the mixing
effect. In this paper, we also particularly emphasized that the
fine-tuning problem associated with the light Higgs mass can be
remarkably mitigated by taking low enough messenger scale
($\approx 300$ GeV) and light enough mass parameters ($\ll 1$
TeV). We have explored the least tuning condition
($\mu_N\lesssim\widetilde{m}_N$), under which even the soft parameters much lighter than 500 GeV can explain the Higgs mass of 125 GeV without conflicting with the LHC experimental results. It is possible because the newly introduced particles are MSSM singlets.

Under the assumption that the observed excess of the diphoton
decay rate of the Higgs over the SM expectation will persist, we
also studied the way to enhance the diphoton decay rate in our
framework. It turns out to be simply realized, only if the hidden
sector fields in our model are converted to carry also electromagnetic charges.
Thus, the mechanism of the Higgs mass enhancement and mitigating
the fine-tuning can be closely related to the excess of the
diphoton decay rate of the Higgs in our framework.

\acknowledgments

The authors thank Kyu Jung Bae and Chang Sub Shin for valuable discussions.
This research is supported by Basic
Science Research Program through the National Research Foundation
of Korea (NRF) funded by the Ministry of Education, Science and
Technology (Grant No. 2010-0009021), and also by Korea Institute
for Advanced Study (KIAS) grant funded by the Korea government
(MEST).


\end{document}